\documentclass[aps, prb, twocolumn, superscriptaddress, amsmath,  tightenlines, longbibliography]{revtex4-1}

\usepackage{dcolumn}
\usepackage{graphicx}
\usepackage{mathrsfs}
\usepackage{subfigure}
\usepackage{booktabs}
\usepackage{amsmath}
\usepackage{physics}
\usepackage{dsfont}
\usepackage{amstext}
\usepackage{amssymb}
\usepackage{amsbsy}
\usepackage{bbm}
\usepackage{amsthm}
\usepackage{graphicx}
\usepackage{color}
\usepackage[colorlinks,citecolor=blue]{hyperref}

\usepackage{url}
\usepackage[colorlinks]{hyperref}
\hypersetup{%
	plainpages=true,
	breaklinks=true,       
	hypertexnames=false,  
	pageanchor=true,
	colorlinks=true,
	linkcolor={blue},
	citecolor={red},
	urlcolor={blue},
	anchorcolor={black}
}

 \makeatletter

\newcommand{\Rmnum}[1]{\expandafter\@slowromancap\romannumeral #1@}
\makeatother

\begin{document}

\title{Interaction-Induced Second-Order Skin Effect}
\author{Wen-Zheng Ling}
\thanks{These authors contributed equally}
\affiliation{School of Physics and Optoelectronics, South China University of Technology,  Guangzhou 510640, China}
\author{Zhao-Fan Cai}
\thanks{These authors contributed equally}
\affiliation{School of Physics and Optoelectronics, South China University of Technology,  Guangzhou 510640, China}
\author{Tao Liu}
\email[E-mail: ]{liutao0716@scut.edu.cn}
\affiliation{School of Physics and Optoelectronics, South China University of Technology,  Guangzhou 510640, China}

\date{{\small \today}}


\begin{abstract}
	In contrast to the conventional (first-order) non-Hermitian skin effect (NHSE) in a $d$-dimensional system with linear size $L$, the $n$th-order (higher-order) NHSE is characterized by skin modes localized at lower-dimensional boundaries of dimension $(d-n)$.  The total number of these modes scales linearly with the system size $L$. Significant progress has been made in understanding higher-order NHSE in non-interacting systems. In this work, we demonstrate the many-body interaction induced second-order skin effect in a two-dimensional non-Hermitian bosonic system. Specifically, we construct a   square lattice that incorporates  nonreciprocal single-boson hopping,  onsite many-body interactions and two-boson pairing hopping. In the absence of interactions, no second-order NHSE is observed. However, with the inclusion of interactions, we identify interaction-induced skin modes for in-gap doublon states (i.e., bound pairs of bosons) localized at the corners of the lattice, while the bulk doublon states remain extended. These corner-localized skin modes arise from the interplay between interaction-induced edge states, localized along one-dimensional boundaries, and the nonreciprocal hopping along these boundaries. Furthermore, the number of corner skin modes scales linearly with the system size, confirming the presence of second-order NHSE in this interacting system. Our findings introduce a novel approach to realizing higher-order skin effects by leveraging interactions.
\end{abstract}

\maketitle

\section{Introduction}\label{section1}

Non-Hermitian systems, governed by Hamiltonians that encapsulate non-conservative dynamics such as particle gain or loss, have emerged as a fertile ground for uncovering unconventional physical phenomena \cite{Ashida2020}. These systems fundamentally challenge conventional notions of symmetry, topology, and dynamics, leading to a surge in both theoretical and experimental investigations in recent years \cite{RevModPhys.93.015005}. Unlike their Hermitian counterparts, non-Hermitian systems exhibit unique behaviors, including exceptional points \cite{Peng2014a, Peng2014b, Gao2015, El-Ganainy2018}, where eigenvalues and eigenvectors coalesce, and non-Hermitian topological phases \cite{PhysRevLett.118.040401, PhysRevLett.118.045701, arXiv:1802.07964,ShunyuYao2018}, which arise from the intricate structure of complex energy bands.  These phenomena underscore the rich and unconventional physics of non-Hermitian systems \cite{PhysRevLett.123.206404, PhysRevLett.123.066405,  ZhangJ2018, Bliokh2019, PhysRevB.100.054105,  PhysRevA.100.062131, Zhao2019, Bliokh2019, PhysRevX.9.041015, PhysRevLett.124.056802, Zhang2022, PhysRevLett.131.036402, PhysRevA.101.062112, PhysRevA.102.033715, PhysRevLett.124.086801, PhysRevLett.127.196801,   Zhang2022,PhysRevLett.130.157201, PhysRevLett.129.093001, Ren2022, PhysRevX.13.021007, PhysRevLett.131.036402,  Leefmans2022, Okuma2023, arXiv:2403.07459, arXiv:2311.03777, PhysRevLett.132.070402, arXiv:2408.12451,PhysRevA.109.063329, Parto2023, PhysRevX.14.021011, PhysRevLett.132.050402, Leefmans2024, Hu2024,Guo2024,Observation2025}, providing profound insights that extend beyond the framework of traditional Hermitian physics.

Non-Hermitian skin effect (NHSE)  is a particularly striking phenomenon, where a macroscopic number of eigenstates are localized at the system boundaries, characterized  by the point-gap topology in the complex energy plane. The NHSE has garnered significant attention in both theoretical and experimental research \cite{ShunyuYao2018, PhysRevLett.125.126402, PhysRevLett.123.066404, YaoarXiv:1804.04672, PhysRevLett.121.026808, PhysRevLett.122.076801, PhysRevLett.123.170401}.  It breaks the conventional bulk-boundary correspondence established by Bloch theory \cite{ShunyuYao2018}, revealing unique physics inherent to non-Hermitian systems. To address this breakdown, a non-Bloch framework based on the generalized Brillouin zone has been proposed \cite{ShunyuYao2018, PhysRevLett.125.126402, PhysRevLett.123.066404, YaoarXiv:1804.04672}, providing a consistent approach to restore the bulk-boundary correspondence in non-Hermitian topological systems. This theoretical advancement has profoundly expanded our understanding of non-Hermitian physics and its interplay with topology \cite{RevModPhys.93.015005}.

In a conventional NHSE for a $d$-dimensional system with linear size $L$, the skin modes are localized at the $(d-1)$ or lower dimensional boundaries, with their total number scaling as the system's volume  $L^d$. This phenomenon is referred to as the first-order NHSE \cite{PhysRevB.102.205118}. Recent studies have generalized this concept to higher-order NHSE (i.e., $n$th-order NHSE) \cite{PhysRevB.102.205118, PhysRevB.102.241202, PhysRevB.103.045420, PhysRevLett.123.016805, PhysRevB.106.035425, Zou2021, PhysRevB.108.075122, PhysRevLett.128.223903, PhysRevLett.131.116601, PhysRevLett.132.063804, Jiang2024, Zhu2024}, where skin modes are localized at lower-dimensional boundaries of dimension $(d-n)$. Notably, in contrast to the first-order NHSE, the number of skin modes in higher-order NHSEs grows with the system's linear size $L$, highlighting a distinct scaling behavior.

Despite significant progress in understanding higher-order NHSE, most studies have concentrated on non-interacting systems, leaving the impact of many-body interactions largely unexplored. Many-body interactions are well-known to profoundly reshape topological phases or even give rise to entirely new phenomena \cite{PhysRevLett.94.086803, PhysRevA.76.023613, Chen2012, PhysRevA.102.013510, PhysRevResearch.2.013348, Olekhno2020, PhysRevLett.133.176601, PhysRevResearch.6.033205}. Incorporating these interactions into non-Hermitian systems opens exciting opportunities for uncovering novel physics and expanding our understanding of non-Hermitian phenomena \cite{Ashida2017, PhysRevLett.121.203001, PhysRevA.102.023306, PhysRevLett.123.123601, PhysRevLett.124.147203, PhysRevB.104.075106, PhysRevB.106.L121102, PhysRevB.106.235125, PhysRevB.105.165137, PhysRevB.102.235151, PhysRevB.106.205147, Shen2022, PhysRevLett.129.180401, PhysRevResearch.5.033173, Longhi2023, PhysRevLett.132.096501, arXiv:2309.14111, arXiv:2405.1228, arXiv:2403.10449, Kim2024, PhysRevResearch.6.L032067}. In interacting non-Hermitian systems, particularly those with nonreciprocal hopping, many-body interactions can generate rich and unconventional phenomena. For instance, interactions can drive a first-order NHSE \cite{PhysRevLett.129.180401}, induce occupation-dependent NHSE \cite{PhysRevLett.132.096501}, and give rise to the non-Hermitian Mott skin effect \cite{arXiv:2309.14111}. These effects highlight the intricate interplay between interactions and non-Hermitian physics, suggesting that many-body correlations can significantly modify or even create new types of non-Hermitian skin effects. This naturally raises a profound question: Can many-body interactions be harnessed to induce higher-order NHSE?

In this work, we explore how the interplay between many-body interactions and the nonreciprocal hopping leads to the emergence of second-order skin effects. Specifically, we study an extended Bose-Hubbard model on a   square lattice, incorporating nonreciprocal single-boson hopping, onsite many-body interactions and two-boson pairing hopping. In the absence of interactions, the system does not exhibit a second-order skin effect. However, when interactions are introduced, two doublon bands (representing bound pairs of bosons) emerge and are separated by a band gap. When the boundary is open along a single direction, interaction-induced in-gap edge states of doublons appear, localized along the one-dimensional boundary. These edge states exhibit a nontrivial point-gap topology, characterized by a nonzero winding number in the complex energy plane. When boundaries are open along both directions, the nonreciprocal hopping along the boundary drives these edge states to localize at the corners of the lattice due to the nontrivial point-gap topology of the one-dimensional edge states. Moreover, the number of corner-localized skin modes scales linearly with the lattice size, providing clear evidence of an interaction-induced second-order NHSE. Importantly, these corner skin modes are robust against disorder. To bridge theory and practice, we propose an experimental setup to realize and observe these predicted phenomena.

The structure of this paper is as follows. In Sec.~\ref{section2}, we introduce a two-dimensional   square lattice model featuring nonreciprocal single-boson hopping, onsite many-body interactions and two-boson pairing hopping. In Sec.\ref{section3}, we investigate the emergence of second-order skin effects and derive an effective doublon Hamiltonian to elucidate the underlying mechanism. The effects of Tamm-Shockley states and disorder are also discussed. Finally, we summarize our findings and conclude the paper in Sec.~\ref{section7}.

\begin{figure}[!tb]
	\centering
	\includegraphics[width=8.7cm]{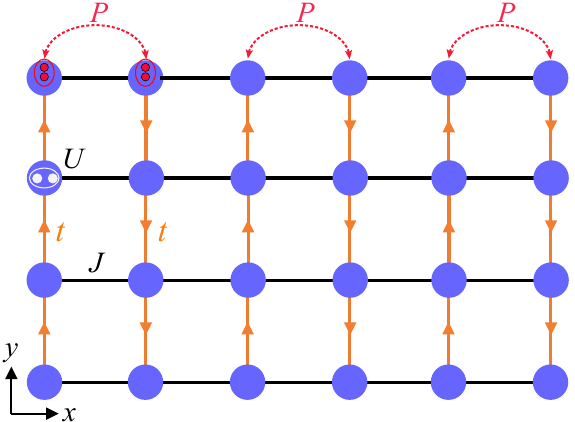}
	\caption{Schematic of the   square lattice model in the presence of many-body interaction. $U$ is the strength of onsite boson-boson interactions. Black solid lines denote the reciprocal single-boson  hopping along the $x$-direction with strength $J$, orange solid  lines with arrows represent the unidirectional single-boson  hopping along the $y$-direction with strength $t$, and red dashed  lines with arrows indicate the reciprocal two-boson pairing hopping along the $x$-direction with strength $P$.  }\label{Fig1}
\end{figure}

\begin{figure*}[!bt]
	\centering
	\includegraphics[width=18.6cm]{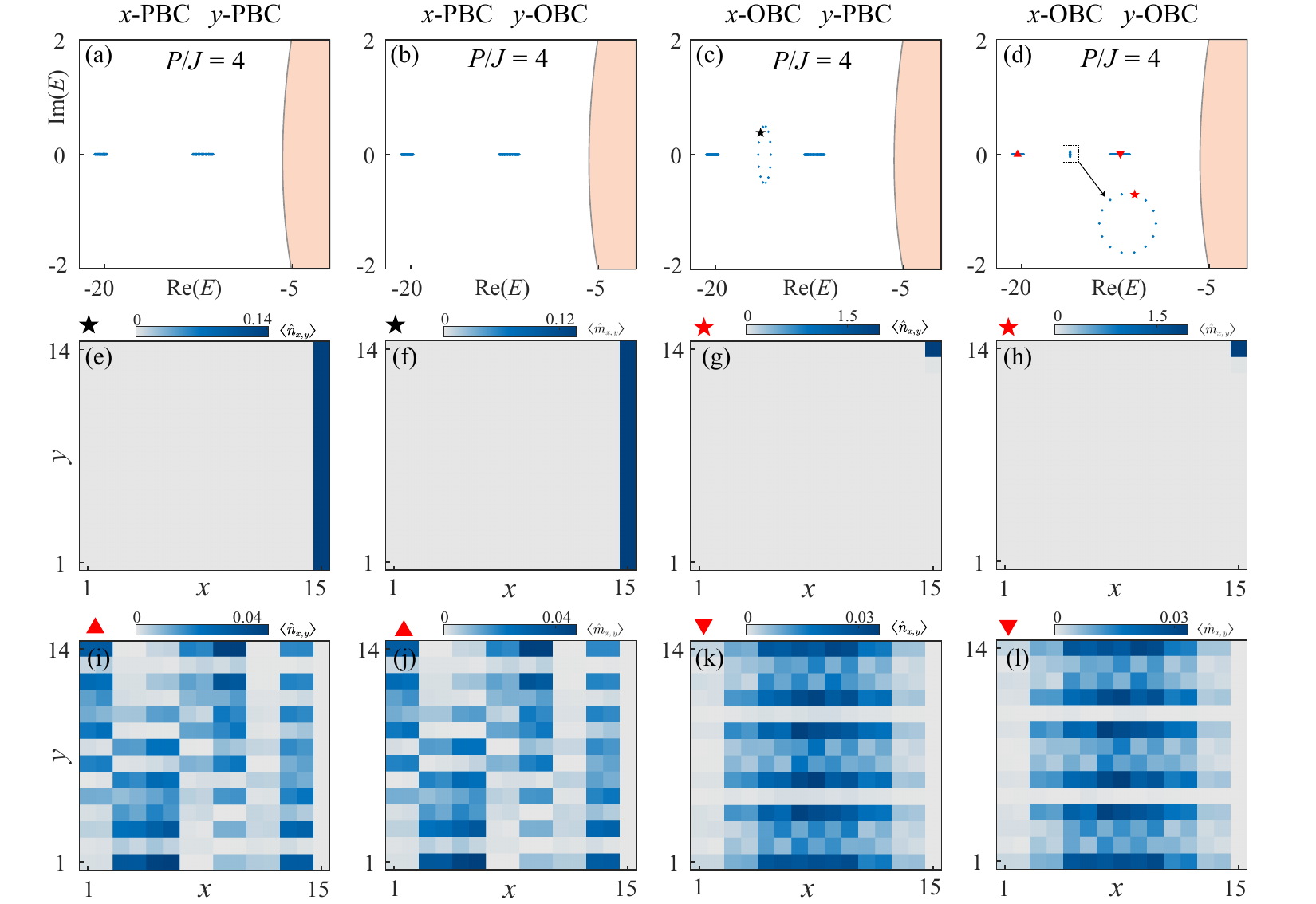}
	\caption{(a-d) Complex eigenenergy spectrum of $\mathcal{H}$  for  $P/J=4$ under various boundary conditions. The OBCs are imported along none of the directions in (a), only along the $y$-direction in (b), only along the $x$-direction in (c), and along both  directions in (d). Apricot-shaded areas represent the continuum of scattering states, and blue dots denote doublon states. The enlarged view of the dotted square region is highlighted by the arrows in (d). (e-l) The corresponding site-resolved single-occupation boson density $\langle \hat{n}_{x,y} \rangle$, and  double-occupation boson density $\langle \hat{m}_{x,y} \rangle $ of the selected energies, respectively. The selected energies are marked by the black and red identifiers within the eigenenergy spectrum in (c) and (d). The lattice size  is set as $L_x \times L_y=15 \times 14$   for the open boundary along the $x$-direction, and for the other boundary conditions, it is $L_x \times L_y = 14 \times 14$. Other parameters used are  $t/J=2$ and $U/J=8$. }\label{Fig2}
\end{figure*}

\begin{figure}[!bt]
	\centering
	\includegraphics[width=8.7cm]{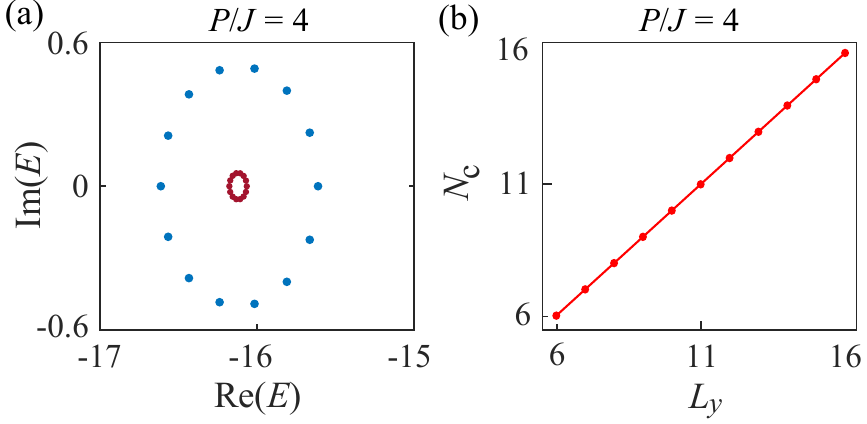}
	\caption{(a) Complex eigenenergy spectrum of in-gap corner skin modes (red dots) and in-gap one-dimensional edge modes (blue dots)  for $P/J=4$. The in-gap corner skin modes are calculated under the OBCs along the $x$ and $y$ directions, and the in-gap edge   modes are calculated under the OBC  along the $x$ direction and the PBC along the $y$ direction.  (b) The number of corner skin modes, $N_c$, as a function of the lattice length $L_y$.   Other parameters are the same as those in Fig.~\ref{Fig2}}\label{Fig4}
\end{figure}

\section{Model}\label{section2}

We study a non-Hermitian extended Bose-Hubbard model on a square lattice, which includes onsite many-body interactions,   two-boson pairing hopping, and unidirectional single-boson hopping, as illustrated in Fig.~\ref{Fig1}. The system Hamiltonian is written as
\begin{align}\label{hamil}
	\mathcal{H}   =  & - J   \sum_{x,y}  \left( \hat{a}_{x+1,y}^\dagger  \hat{a}_{x,y} +\textrm{H.c.} \right)  - U\sum_{x,y} \hat{n}_{x,y} \left(\hat{n}_{x,y}-1 \right) \nonumber \\
	& -\frac{P}{2} \sum_{x,y}\left(\hat{a}_{2x-1,y}^\dagger \hat{a}_{2x-1,y}^\dagger  \hat{a}_{2x,y}  \hat{a}_{2x,y}+ \textrm{H.c.}\right) \nonumber \\
	& -i t \sum_{x,y}  \left(\hat{a}_{2x-1,y+1}^\dagger  \hat{a}_{2x-1,y} + \hat{a}_{2x,y}^\dagger  \hat{a}_{2x,y+1}\right),
\end{align}
where $\hat{a}^\dagger_{x,y}$ and $\hat{a}_{x,y}$ denote bosonic creation and annihilation operators at site $(x,y)$, and $\hat{n}_{x,y} = \hat{a}_{x,y}^\dagger \hat{a}_{x,y}$ is the particle number operator, $J$ and $t$ are the amplitudes of reciprocal  and unidirectional  single-boson hopping along the $x$- and $y$-direction, respectively. The interaction strength $U$ characterizes the local interactions between bosons, and $P$ denotes the amplitude of direct two-boson pairing hopping. The Hamiltonian $\mathcal{H} $ in Eq.~(\ref{hamil}) can be realized by utilizing the  post-selection measurement    \cite{Ashida2020} in e.g., ultracold-atom platforms, as explained in Appendix \ref{AppendixA}. 

The Hamiltonian $\mathcal{H}$ exhibits $U(1)$ symmetry, as evidenced by its commutation with the total particle number operator, $[\mathcal{H}, ~\sum_{x,y} \hat{n}_{x,y}] = 0$. This symmetry ensures that the system can be analyzed within a fixed particle-number subspace of the Fock space. In this work, we specifically focus on the double-excitation subspace and attractive interaction with $U>0$.

\section{Interaction-induced second-order corner skin  modes }\label{section3}

\subsection{In-gap corner skin  modes of doublons}

In the single-boson case, the 2D lattice described in Eq.~(\ref{hamil}) is trivial, which exhibits no second-order skin effects [see details in Appendix \ref{AppendixB2}]. In the double-excitation subspace, the onsite many-body interaction leads to the formation of bound boson pairs (i.e., doublons)  for both repulsive and attractive interaction regimes \cite{Winkler2006}. The interplay between many-body interactions-induced edge states, localized at the one-dimensional boundary, and the nonreciprocal hopping along the boundary gives rise to doublon skin  modes [see detailed discussions in Sec.~III(B) and Appendix \ref{AppendixB}], which are localized at the lattice corner. In contrast, the bulk states remain extended across the bulk sites due to destructive interference caused by the opposing unidirectional hopping along the $y$ directions within each square plaquette. 

We firstly numerically explore the phenomenon of second-order skin effects induced by the many-body interactions. We numerically calculate the complex spectrum and the particle density for single occupation and double occupation defined, respectively, as
\begin{align}\label{particledensityn}
	\langle \hat{n}_{x,y} \rangle = \frac{_R\langle \psi_m | \hat{n}_{x,y} | \psi_m \rangle_R}{_R \langle \psi_m | \psi_m \rangle_R},
\end{align}
\begin{align}\label{particledensitym}
	\langle \hat{m}_{x,y} \rangle = \frac{_R\langle \psi_m | \hat{m}_{x,y} | \psi_m \rangle_R}{_R \langle \psi_m | \psi_m \rangle_R},
\end{align}
where $\hat{m}_{x,y}=\hat{a}^\dagger_{x,y}\hat{a}^\dagger_{x,y}\hat{a}_{x,y}\hat{a}_{x,y}$ and $\ket{\psi_m}_R (m=1,2,3,\cdots)$ is the $m$-th right eigenvector of the non-Hermitian interacting Hamiltonian satisfying $\mathcal{H} \ket{\psi_m}_R = E_m \ket{\psi_m}_R$.

Figure \ref{Fig2} shows	the complex eigenspectrum  $E$ and the corresponding site-resolved single-occupation boson density $\langle \hat{n}_{x,y} \rangle$, and  double-occupation boson density $\langle \hat{m}_{x,y} \rangle $   under different boundary conditions for $P/J=4$.  The apricot-shaded regions represent the scattering states, and the blue dots correspond to doublon states. The eigenspectrum is distinctly partitioned into a continuum of scattering states and discrete doublon bands. The scattering states correspond to superposition of two-particle configurations with particles localized on different lattice sites. In contrast, the doublon bands are composed of bound boson states, where both particles are co-localized on the same site. In addition, the eigenenergies of the doublon bands are well separated from the continuum of scattering states.

There is a band gap between two doublon bands under the periodic boundary conditions (PBCs) along both the $x$ and $y$ directions, as shown in Fig.~\ref{Fig2}(a). When the boundary condition  only along  the $y$ direction is open, no in-gap modes are observed [see Fig.~\ref{Fig2}(b)]. In contrast, when the boundary is open only along the $x$ direction, in-gap modes emerge [see Figs.~\ref{Fig2}(c)]. These in-gap modes are localized at the right edge, as illustrated in Fig.~\ref{Fig2}(e,f), which can be understood as topological edge states localized at  weakly-linked edge sites [see details in  Sec.~III(B)]. By comparing the distributions of the single-occupation density $\langle \hat{n}_{x,y} \rangle$ [see Fig.~\ref{Fig2}(e)] and  the double-occupation density $\langle \hat{m}_{x,y} \rangle$ [see Fig.~\ref{Fig2}(f)], we can deduce the characteristics of the bound-boson pair for each in-gap mode.

In the case of OBCs along both the $x$ and $y$ directions, in-gap modes still persist [see Fig.~\ref{Fig2}(d)]. Due to the unidirectional hopping along the $y$ direction in the right boundary, these modes become localized at the top-right corners, as reflected in the single-occupation density   $\langle \hat{n}_{x,y} \rangle$ [see Fig.~\ref{Fig2}(g)] and  the double-occupation density $\langle \hat{m}_{x,y} \rangle$ [see Fig.~\ref{Fig2}(h)]. Importantly, these in-gap corner modes exhibit the characteristics of bound boson pairs. Moreover, the corner-localized states under OBCs in both directions can be interpreted as the skin modes of the one-dimensional edge states that arise when OBCs are imposed only along the $x$ direction. In this scenario, the eigenspectrum of the former case is enclosed by the point-gap loop of the latter case [see Fig.~\ref{Fig4}(a)]. This corner skin effect originates from the intrinsic non-Hermitian topology of the one-dimensional edge states. Furthermore, the number of corner skin modes scales with the linear system size [see Fig.~\ref{Fig4}(b)],  as inherited from the edge modes. Note that in the case of a first-order skin effect, the number of skin modes scales with $L_x \times L_y$, regardless of variations in $L_y$.   This
is the hallmark of the second-order skin effect \cite{PhysRevB.102.205118}.

In order to characterize this point-gap topology, we introduce a twisted boundary, with twist angle $\varphi$ ($\varphi \in [0,~2\pi]$), along the $y$ direction and OBC along the $x$  direction. We then   define the many-body winding number \cite{PhysRevB.106.L121102,PhysRevB.105.165137,PhysRevB.106.205147}
\begin{align}\label{winding}
	\mathcal{W} = \oint_{0}^{2\pi} \frac{d \varphi}{2\pi i} \frac{\partial}{\partial \varphi} \log\left[\det \left(\hat{\mathcal{H}}(\varphi) -E_\textrm{ref}\right)\right], 
\end{align}
where $E_\textrm{ref}$ is the complex reference energy inside the point gap of the in-gap one-dimensional edge states [see blue dots in Figs.~\ref{Fig4}(a)] . The nonzero winding number  with $\mathcal{W}=2$  indicates the intrinsic topological   origin of corner skin modes.  These   results show that  the interplay between many-body interactions-induced edge states and the nonreciprocal hopping along the boundary results in the formation of doublon skin modes localized at the lattice corner [see further details in Sec.~III(B)]. In contrast, the bulk states of doublons remain extended across the bulk sites due to destructive interference caused by the opposing unidirectional hopping along the $y$ directions within each square plaquette [see density distributions of bulk bands in Fig.~\ref{Fig2}(i-l)].

\subsection{Effective Hamiltonian of doublons in strong-interaction limit}\label{effeHamil}
 	
We numerically demonstrate the occurrence of interaction-induced corner skin modes of doublons within the double-excitation subspace in Sec.~III(A). To gain deeper insight into the underlying two-body interaction mechanism, we derive an effective Hamiltonian for doublons in the strong-interaction limit  with $\abs{U} \gg ~\abs{J}, ~\abs{t},~\abs{P}$. 

We rewrite the Hamiltonian in Eq.~(\ref{hamil}) as $\mathcal{H} = \mathcal{H}_\textrm{0} + \mathcal{V}$, where $\mathcal{H}_\textrm{0}$ is the dominant interaction part with
\begin{align}\label{hamil0}
	\mathcal{H}_\textrm{0}   = - U \sum_{x,y} \hat{n}_{x,y} \left(\hat{n}_{x,y}-1 \right) ,
\end{align}
and $\mathcal{V}$ is perturbed  hopping terms  in the strong-interaction limit with 
\begin{align}\label{VV}
	\mathcal{V}   = 	&-i t \sum_{x,y}  \left(\hat{a}_{2x-1,y+1}^\dagger  \hat{a}_{2x-1,y} + \hat{a}_{2x,y}^\dagger  \hat{a}_{2x,y+1}\right) \notag \nonumber \\
	&-\frac{P}{2} \sum_{x,y} \left(\hat{a}_{2x-1,y}^\dagger \hat{a}_{2x-1,y}^\dagger  \hat{a}_{2x,y}  \hat{a}_{2x,y} +  \textrm{H.c.}\right) \notag \nonumber \\ 
	&  - J   \sum_{x,y}  \left( \hat{a}_{x+1,y}^\dagger  \hat{a}_{x,y} +\textrm{H.c.} \right).
\end{align}

In the strong-interaction limit  with $\abs{U} \gg ~\abs{J}, ~\abs{t},~\abs{P}$, two bosons are tightly bound to form a bound state (doublon), such that the set of doublon states $|d_{x,y}\rangle = \hat{d}^\dagger_{x,y} |0\rangle= \hat{a}^\dagger_{x,y} \hat{a}^\dagger_{x,y} |0\rangle$ are eigenstates of the unperturbed Hamiltonian $\mathcal{H}_\textrm{0}$ with energy $E_d = -2U$. All the other two-boson scattering states are of form $|s\rangle = \hat{a}^\dagger_{x',y'} \hat{a}^\dagger_{x,y} |0\rangle$, where $x'\neq x,~y'\neq y$, and have energy $E_s = 0$. For convenience, we adopt the shorthand notation $\ket{d}\equiv\ket{d_{x,y}}$. By utilizing quasi-degenerate second-order perturbation theory \cite{Bir1974,CCohenTannoudji1Atom}, the nonzero matrix elements of the effective Hamiltonian $\mathcal{H}_\textrm{eff}$ are given by 
\begin{align}\label{perturbation22}
	\langle d|\mathcal{H}_\textrm{eff}|d'\rangle = & ~E_d \delta_{d,d'} + \langle d |\mathcal{V}|d'\rangle \nonumber \\
	&+  \frac{1}{2} \sum_{s} \langle d|\mathcal{V}|s\rangle \langle s |\mathcal{V}|d'\rangle \times \left(\frac{2}{E_d-E_s}\right),
\end{align}
where $E_n$ is the $n$th eigenvalue  of $\mathcal{H}_\textrm{0}$ relative to eigenstate $|n\rangle$. 

According to Eq.~(\ref{perturbation22}),  as shown in Appendix  \ref{AppendixB},  we obtain the effective Hamiltonian $\mathcal{H}_\textrm{eff}$ acted on quasi-single-particle subspace as
 	\begin{align}\label{hamileffappendix2}
 		\mathcal{H}_\textrm{eff} = & - U_\textrm{eff} \sum_{x,y} \hat{d}^\dagger_{x,y} \hat{d}_{x,y} - J_\text{eff} \sum_{x,y} \left( \hat{d}^\dagger_{2x+1,y} \hat{d}_{2x,y} + \text{H.c.}\right) \notag \nonumber \\
 		&- (J_\text{eff}+P) \sum_{x,y} \left(\hat{d}^\dagger_{2x,y} \hat{d}_{2x-1,y} + \text{H.c.}\right) \notag \nonumber \\
 		&+ t_\text{eff} \sum_{x,y} \left(\hat{d}^\dagger_{2x-1,y+1} \hat{d}_{2x-1,y} + \hat{d}^\dagger_{2x,y} \hat{d}_{2x,y+1}\right),
 	\end{align}
 where $J_\textrm{eff} = J^2/U$ is the effective symmetric doublon hopping along the $x$ direction, and $t_\textrm{eff} = t^2/U$ is the effective unidirectional along the $y$ direction. For the PBC, the effective onsite energy is given by $U_\textrm{eff} = 2J^2/U+2U$. When the boundary along the $x$ direction is open, the onsite energy reads $U_\textrm{eff} = 2J^2/U+2U$ for the bulk sites, and $U_\textrm{eff} = J^2/U+2U$ for sites located at the left and right edges.

 \begin{figure}[!t]
 	\centering
 	\includegraphics[width=8.7cm]{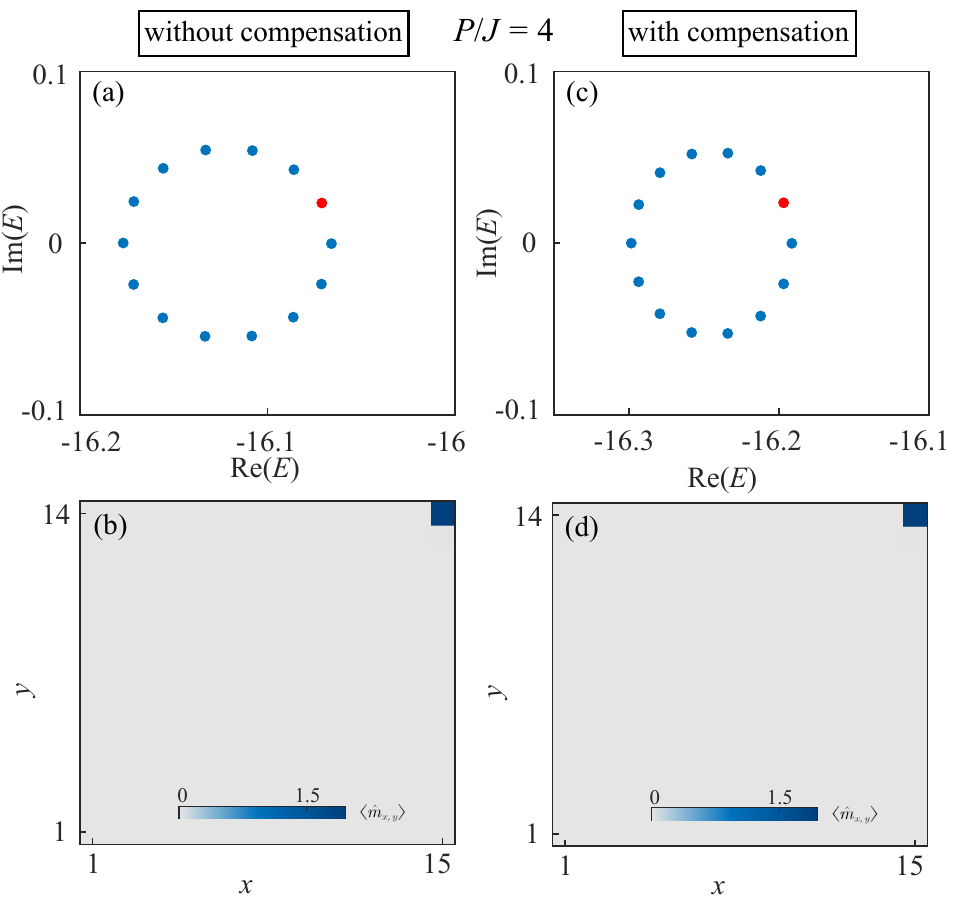}
 	\caption{Complex eigenenergy spectrum of in-gap corner skin modes  states (a)  without the compensating potential $V$, and (c) with the potential $V = J^2/U$ applied   to both left and right edges to compensate for the interaction-induced detuning.  The corresponding site-resolved double-occupation density $\langle \hat{m}_{x,y} \rangle$ for  the selected in-gap state [see red markers in (a,c)] is plotted in (b) and (d), respectively. The parameters are the same as those in Fig.~\ref{Fig2}. }\label{Fig5}
 \end{figure}

According to the effective Hamiltonian $\mathcal{H}_\textrm{eff}$ in Eq.~(\ref{hamileffappendix2}),  a doublon behaves as a quasiparticle in a 2D square lattice under the strong-interaction limit. Along the $y$ direction,  the quasiparticle undergoes unidirectional hopping, with its motion alternating between upward and downward directions. In the absence of hopping along the $y$ direction, the model in Eq.~(\ref{hamileffappendix2}) decouples into identical Su–Schrieffer–Heeger (SSH) chains with two alternating tunneling amplitudes $J_\text{eff}+P$ and $J_\text{eff}$ along the $x$ direction. Under OBC in the $x$ direction for each SSH chain, localized edge modes emerge.  However, unlike the standard SSH model \cite{PhysRevLett.42.1698}, this decoupled  SSH  chain  features edge sites with onsite energies detuned by $J^2/U$   relative to the bulk sites. 
 
Despite this detuning, previous studies  been shown that when each pair-hopping chain in the $x$ direction is decoupled along the $y$ direction, it supports topologically protected weakly-linked boundary states localized at the right edge for odd sites in the $x$ direction \cite{PhysRevA.102.013510,Olekhno2020}. Here, we provide a detailed analysis of the existence of these topologically protected weakly linked boundary states for the decoupled pair-hopping chain along the $x$ direction  by employing the effective Hamiltonian $\mathcal{H}_\textrm{eff}$ in Eq.~(\ref{hamileffappendix2}).

For the decoupled pair-hopping chain along the $x$ direction under OBCs, using Eq.~(\ref{hamileffappendix2}) in the strong-interaction limit, the effective Hamiltonian reads  
\begin{align}\label{hamileffappendix2inter}
	\mathcal{H}_\textrm{1D} = & - U_\textrm{eff} \sum_{x} \hat{d}^\dagger_{x} \hat{d}_{x} - J_\text{eff} \sum_{x} \left( \hat{d}^\dagger_{2x+1} \hat{d}_{2x} + \text{H.c.}\right) \notag \nonumber \\
	&- (J_\text{eff}+P) \sum_{x} \left(\hat{d}^\dagger_{2x} \hat{d}_{2x-1} + \text{H.c.}\right),
\end{align}
where the onsite energy reads $U_\textrm{eff} = 2J^2/U+2U$ for the bulk sites, and $U_\textrm{eff} = J^2/U+2U$ for sites located at the left and right edges.

 By solving the Schr\"{o}dinger equation $\mathcal{H}_\textrm{1D} \ket{\psi} = \xi_0 \ket{\psi} $, with $\ket{\psi}  =    \sum_{x} \beta_{x} \hat{d}^\dagger_{x} \ket{x}$, we have 

 	\begin{align}\label{eqsbeta22}
 		-\frac{J^2}{U}  \beta_{2 x-2}  - \left(P + \frac{J^2}{U} \right) \beta_{2 x} =  \left(\xi_0 + 2 U + \frac{2J^2}{U} \right) \beta_{2 x-1 },   
 	\end{align}

 	\begin{align}\label{eqsbeta32}
 		-\frac{J^2}{U}  \beta_{2 x+1 } - \left(P + \frac{J^2}{U} \right)\beta_{2 x-1 } =   \left(\xi_0 + 2 U + \frac{2J^2}{U} \right) \beta_{ 2 x}.  
 	\end{align}

To be particular, the coefficients $\beta_{1}$ and $\beta_{N_x}$ ($N_x$ is length) at both edges of the decoupled pair-hopping chain satisfy
 \begin{align}\label{edge222}
 	- \left(P + \frac{J^2}{U} \right) \beta_{2} =  \left(\xi_0 + 2 U + \frac{J^2}{U} \right) \beta_{1},   
 \end{align}
 \begin{align}\label{eqsbeta322}
 	-\frac{J^2}{U}  \beta_{L_x-1}   =  \left(\xi_0 + 2 U + \frac{J^2}{U} \right) \beta_{L_x}, 
 \end{align}
 where $L_x$ is assumed to be odd.
 
According to Eqs.~(\ref{eqsbeta22}) and (\ref{eqsbeta32}), in the strong-interaction limit, the doublons in the decoupled pair-hopping chain along the $x$-direction behave as quasiparticles in an unconventional SSH model, with the detuned boundary sites acting as defects. These defects can give rise to topologically nontrivial boundary states \cite{PhysRevA.102.013510,Olekhno2020}.
 
We now find the boundary-localized solutions. Firstly, in order to solve Eqs.~(\ref{eqsbeta22}) and (\ref{eqsbeta32}) under the boundary condition (\ref{eqsbeta322}) for the boundary state localized at the right boundary, we choose trial solutions $\beta_{2x} = \zeta^{2x} \varphi_B$ and $\beta_{2x-1 } = \zeta^{2x-1} \varphi_A$ with $\abs{\zeta}>1$. Then, we have 
 \begin{align}\label{edge1}
 	\left(\xi_0 + 2 U + \frac{2J^2}{U} \right) \varphi_A  = \left[-\frac{J^2}{U}  \zeta^{-1}     - \left(P + \frac{J^2}{U} \right) \zeta  \right]  \varphi_B  ,   
 \end{align}
 \begin{align}\label{edge2}
 	\left(\xi_0 + 2 U + \frac{2J^2}{U} \right) \varphi_B	= \left[-\frac{J^2}{U}  \zeta  - \left(P + \frac{J^2}{U} \right)  \zeta^{-1} \right]    \varphi_A,  
 \end{align} 
 \begin{align}\label{edge3}
 	-\frac{J^2}{U}  \varphi_B   =  \zeta \left(\xi_0 + 2 U + \frac{J^2}{U} \right) \varphi_A, 
 \end{align}
According to Eqs.~(\ref{edge1}-\ref{edge3}), we have two states with   their energies $\epsilon_{\pm}$ and $\zeta^2_\pm$ as 
 \begin{align}\label{localizationenerg}
 	\epsilon_{\pm} = -J_0 - 2U-\frac{2 J_0 P + P^2 \pm \sqrt{(2 J_0 P + P^2)^2+4 J_0^4}}{2 J_0}, 
 \end{align}
 \begin{align}\label{localization}
 	\abs{\zeta^2_\pm} = \abs{\frac{2 J_0 P + P^2 \pm \sqrt{(2 J_0 P + P^2)^2+4 J_0^4}}{2 J_0 (J_0 + P)}},
 \end{align}
 where $J_0 = J^2/U$. Obviously, for any $P \neq 0$, the lower-energy boundary state $\epsilon_+$ is localized.   For the higher-energy state $\epsilon_-$, it is disappears when
 \begin{align}\label{condition}
 	P>0 ~~~ \textrm{or} ~~~ P < -2J^2/U.
 \end{align}
Therefore, with condition in Eq.~(\ref{condition}), the pair-hopping chain supports the edge states localized at the weakly-linked edge near the site $L_x$ with $\abs{P+ J^2/U} >\abs{ J^2/U}$. This edge state  can be intuitively
 interpreted as the topological state inherent to SSH model \cite{PhysRevA.102.013510,Olekhno2020}.
 
Secondly,   in order to solve Eqs.~(\ref{eqsbeta22}) and (\ref{eqsbeta32}) under the boundary condition (\ref{edge222}) for the edge localized at the left boundary, we choose trial solutions $\beta_{2x} = \zeta^{2x} \varphi_B$ and $\beta_{2x-1 } = \zeta^{2x-1} \varphi_A$ with $\abs{\zeta}<1$. Using the same derivation with the condition in Eq.~(\ref{condition}), we cannot find  modes  localized at the left boundary.

I have demonstrated that the  decoupled pair-hopping chain along the $x$-direction in the strong-interaction limit supports the weakly-linked boundary state localized at the right edge for $P>0$ or  $P < -2J^2/U$. When the nonreciprocal hopping along the $y$ direction are turned on, the bulk states remain extended across the bulk sites due to destructive interference caused by the opposing unidirectional hopping along the $y$ directions within each square plaquette. However, the hopping along the $y$ direction remains locally unidirectional at the right boundary,  edge states are localized towards the top-right corner. Therefore, the interplay between many-body interactions-induced one-dimensional edge states and the unidirectional hopping along this edge results in the formation of doublon skin modes localized at the lattice corner. Note that we have shown the appearance of corner skin modes for $P>0$, the results for $P < -2J^2/U$ are presented in Appendix \ref{AppendixD}.

 \begin{figure}
 	\centering
 	\includegraphics[width=8.7cm]{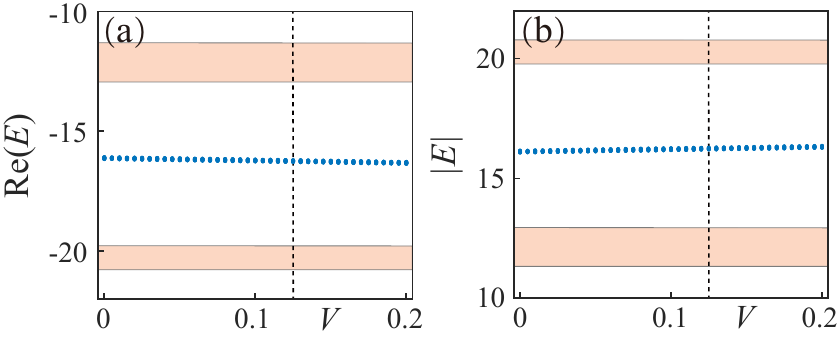}
 	\caption{(a) Real part and (b) absolute value of eigenenergies of doublon states as a function of the compensating potential $V$ applied to both left and right edges to compensate the interaction-induced detuning. The apricot-shaded areas represent the  bulk states of two doublon bands, while the blue dots indicate the in-gap doublon states. The dashed black lines correspond to the compensated potential $V=J^2/U$. The parameters used are $t/J=2$, $P/J=4$, $U/J=8$,  and $L_x\times L_y=13\times12$.  }\label{Fig6}
 \end{figure}

\subsection{Effects of the Tamm-Shockley states}\label{section4}

As discussed in Sec.~\ref{section3}(B), the onsite energies at the edge sites are shifted by $J^2/U$ relative to the bulk sites due to the onsite many-body interaction under OBCs [see details in Appendix \ref{AppendixB}]. This distinct renormalization of system parameters at the edges compared to the bulk results in the formation of Tamm-Shockley states localized at the left and right edge sites \cite{Shockley1932,PhysRev.56.317}. These states arise from non-topological mechanisms. Furthermore, the unidirectional hopping along the $y$ direction may lead to the emergence of corner states from these Tamm-Shockley edge states. 

To verify that the corner states discussed above are induced by the interplay between the interaction-driven  topological edge states and the nonreciprocal hopping along the edge, rather than by Tamm-Shockley states, we introduce a compensating potential $V$, applied to the left and right edges. This compensating potential adjusts the onsite potential energy of edge sites to $-2U - J^2/U - V$,  thereby eliminating the effects of the distinct renormalization of system parameters at the edges relative to the bulk due to the onsite many-body interaction.

Figures  \ref{Fig5}(a) and \ref{Fig5}(c) present  the comparative complex eigenenergy spectra of in-gap corner skin states for $P/J=4$,  without and with the application of a compensating potential $V=J^2/U$. This potential $V=J^2/U$ is applied to both the left and right edges to counteract the interaction-induced detuning.  The corresponding site-resolved single-occupation density $\langle \hat{n}_{x,y} \rangle$ for all the in-gap states are shown in Fig.~\ref{Fig5}(b,d). After applying the compensated potential, the in-gap doublon states remain well localized at the corners, despite a slight deviation in eigenenergy. This indicates that the corner skin modes do not arise from Tamm-Shockley edge states but are instead induced by the interplay between the interaction-driven  topological edge states and the nonreciprocal hopping along the edge, as discussed above. Moreover, we calculate the complex eigenenergy spectrum of doublon states as a function of the compensating potential $V$, as shown in Fig.~\ref{Fig6}, which provides further evidence. As $V$ increases, the corner skin modes (blue dots) remain distinct and do not merge into the doublon bulk bands (apricot-shaded regions), confirming their topological nature.

 \begin{figure}[!bt]
	\centering
	\includegraphics[width=8.7cm]{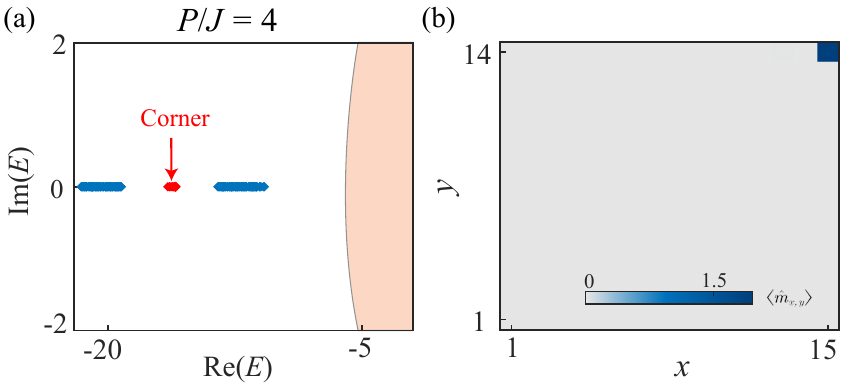}
	\caption{(a) Complex eigenenergy spectrum of $\mathcal{H}$, subject to the random disorder applied to the hopping terms $J$, $t$ and $P$ for  $P/J=4$ and $W/J=2$. (b) The corresponding site-resolved density $\langle \hat{m}_{x,y} \rangle$ of the selected corner state. Other parameters are the same as those  in Fig.~\ref{Fig2}.  }\label{Fig7}
\end{figure}
 
\subsection{Effects of disorder}\label{section5}

To further validate the topological nature of interaction-induced  corner skin modes, we examine the impact of disorder. We introduce the random disorder into the hopping terms $J$, $t$ and $P$ in the Hamiltonian $\mathcal{H}$. The disordered Hamiltonian is written as $\mathcal{H}_\textrm{tot} = \mathcal{H} + \mathcal{H}_\textrm{dis}$, with the disordered term $\mathcal{H}_\textrm{dis}$ being
\begin{align}\label{hamildisorder}
	\mathcal{H}_\textrm{dis}   =  & -     \sum_{x,y}  \tilde{J}_x \left( \hat{a}_{x+1,y}^\dagger  \hat{a}_{x,y} +\textrm{H.c.} \right)   \nonumber \\
	& -  \sum_{x,y} \tilde{P}_{2x-1}\left(\hat{a}_{2x-1,y}^\dagger \hat{a}_{2x-1,y}^\dagger  \hat{a}_{2x,y}  \hat{a}_{2x,y}+ \textrm{H.c.}\right) \nonumber \\
	& -i  \sum_{x,y} \tilde{t}_y \left(\hat{a}_{2x-1,y+1}^\dagger  \hat{a}_{2x-1,y} + \hat{a}_{2x,y}^\dagger  \hat{a}_{2x,y+1}\right),
\end{align}
where  $\tilde{J}_x$, $\tilde{P}_x$, and $\tilde{t}_y$ are randomly drawn from a uniform distribution over the range $[-W/2, ~W/2]$.

We calculate the complex eigenenergy spectrum of $\mathcal{H}$ under random disorder applied to  the hopping terms $J$, $t$ and $P$ for $P/J=4$, as shown in Fig.~\ref{Fig7}(a). Despite the presence of strong disorder, the band gap of the doublon bands and the in-gap states persist. Moreover, these in-gap doublon states remain well localized at the top-right corner of the lattice [see Fig.~\ref{Fig7}(b)], demonstrating their robustness against disorder. The robustness arises from the topological nature of the corner skin states, which is driven by the interplay between the interaction-induced   edge localization and the   nonreciprocal hopping.

\section{Discussion and Conclusion}\label{section7}

We have theoretically demonstrated interaction-induced corner skin modes of doublons (bound pairs of bosons) in a   square lattice featuring nonreciprocal single-boson hopping, onsite many-body interactions, and two-boson pairing hopping. While the single-particle model does not exhibit a second-order skin effect, the two-particle doublon bands feature a band gap in which the in-gap states are localized at the corners. These corner states arise from the interplay between interaction-induced edge states, localized along the one-dimensional boundaries, and the nonreciprocal hopping along these boundaries. Under full open boundaries, the corner skin modes are enclosed by the point-gap loop of the in-gap doublon states in systems with open boundaries along a single direction, characterized by a non-zero winding number. Thus, these corner states can be identified as the skin modes of the edge states, robust against disorder. Furthermore, the number of corner skin modes scales with the lattice size, showcasing the second-order skin effect. Our study provides a novel approach to realizing higher-order skin effects by leveraging many-body interactions. 

Future research directions include extending our model to explore higher-dimensional and higher-order skin effects, as well as systems involving excitation subspace  beyond two. For example, a third-order skin effect may be realized by introducing nonreciprocal single-particle hopping along the $z$-direction in a three-dimensional system. Additionally, in two-dimensional systems, second-order skin effects with corner-localized modes could emerge when considering excitation subspace  beyond two (see Appendix ~\ref{AppendixE} for a simple discussion of the three-excitation case). Another intriguing direction is to investigate interaction-induced higher-order skin effects in systems governed purely by gain and loss.

\begin{acknowledgments}
	T.L. acknowledges support from the National Natural Science Foundation of
	China (Grant No. 12274142), the Introduced Innovative Team Project of Guangdong Pearl River Talents Program (Grant No. 2021ZT09Z109), the Fundamental Research Funds for the Central Universities (Grant No. 2023ZYGXZR020),  and the Startup Grant of South China University of Technology (Grant No. 20210012).
\end{acknowledgments}

\begin{figure*}[!tb]
	\centering
	\includegraphics[width=18cm]{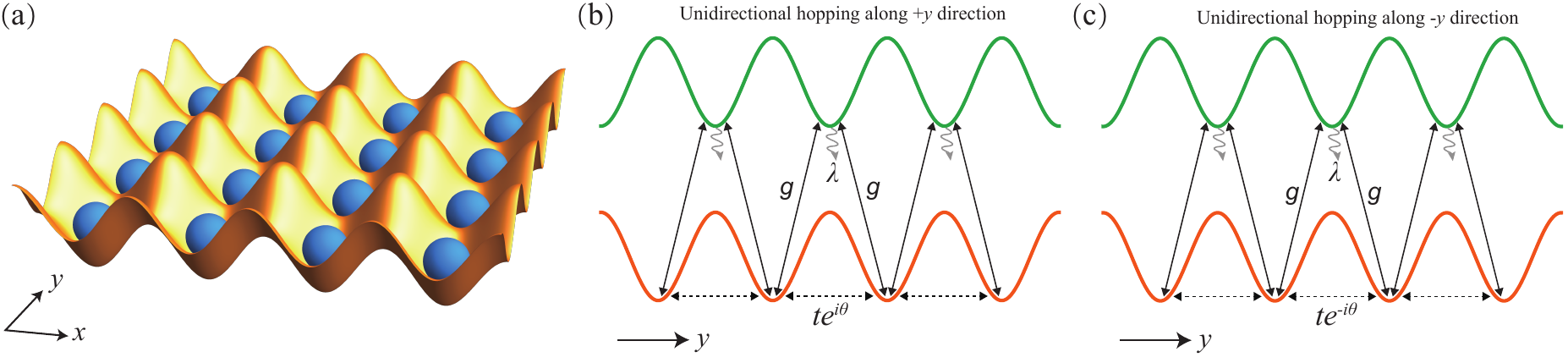}
	\caption{Implementation of asymmetric hopping in optical lattices. (a) Schematic diagram of the experimental setup for a primary optical lattice. The primary lattice (orange curves) is coherently coupled to an auxiliary optical lattice (green curves), where alternating magnetic fluxes  $\theta$ (b) and $-\theta$ (c) are applied along the 
		$x$-direction,  threading through each plaquette along the $y$-direction. The dissipative auxiliary optical lattice, combined with alternating magnetic fluxes, induces unidirectional hopping along the $+y$ and $-y$ directions. This unidirectional hopping alternates along the $x$-direction within the primary optical lattice. }\label{FigS1}
\end{figure*}

\appendix
\section{Experimental Realization}\label{AppendixA}

The model discussed in this paper can be implemented across various platforms, including superconducting qubit array \cite{Roushan2016,PhysRevLett.128.213903}, ultracold atom \cite{PhysRevLett.129.070401, Ren2022,PhysRevLett.122.023601}, and   digital quantum computer   \cite{PhysRevLett.129.140502,arXiv:2311.10143}. Additionally, it can be simulated using electrical circuit \cite{Olekhno2020}. In this Appendix, we present an experimental framework for realizing the interaction-induced second-order skin effect in interacting ultracold atoms loaded into an optical lattice, utilizing two advanced techniques, as shown in Fig.~\ref{FigS1}. The first technique involves the implementation of an artificial gauge field \cite{Goldman2016, Grg2019, Weitenberg2021}, which introduces  magnetic fluxes $ \theta$ and $-\theta$ threading through each plaquette [see Fig.~\ref{FigS1}(b,c)]. The second technique is dissipation engineering \cite{PhysRevLett.77.4728,Mller2012, PhysRevLett.129.070401,PhysRevX.8.031079}, which allows for asymmetric modification of the hopping amplitude along the $y$-direction.

Specifically, we consider two coupled optical lattices: the primary and auxiliary lattices. Figure \ref{FigS1}(a) schematically illustrates the primary optical lattice. The primary lattice (orange curves) is coherently coupled to an auxiliary optical lattice (green curves), where alternating magnetic fluxes  $\theta$ and $-\theta$ are applied along the 
$x$-direction,  threading through each plaquette along the $y$-direction, as shown in Fig.~\ref{FigS1}(b,c).  The dissipative auxiliary optical lattice, combined with alternating magnetic fluxes, induces unidirectional hopping along the $+y$ and $-y$ directions. This unidirectional hopping alternates along the $x$-direction within the primary optical lattice.

The total Hamiltonian of the coupled primary and auxiliary  lattices are written as $\mathcal{H}_\text{total}   =  \mathcal{H}_\textrm{a} + \mathcal{H}_\textrm{b} + \mathcal{H}_\textrm{int}$, where the tight-binding Hamiltonian of the primary lattice is given by
\begin{align}\label{Hamila}
	\mathcal{H}_\textrm{a}   =  & - J   \sum_{x,y}  \left( \hat{a}_{x+1,y}^\dagger  \hat{a}_{x,y} +\textrm{H.c.} \right) - U\sum_{x,y} \hat{n}_{x,y} \left(\hat{n}_{x,y}-1 \right) \nonumber \\
	& - \frac{P}{2} \sum_{x,y} \left(\hat{a}_{2x-1,y}^\dagger \hat{a}_{2x-1,y}^\dagger  \hat{a}_{2x,y}  \hat{a}_{2x,y}+ \textrm{H.c.}\right) \nonumber \\
	& - \sum_{x,y}  \left( \frac{t e^{i\theta}}{2} \hat{a}_{2x-1,y+1}^\dagger  \hat{a}_{2x-1,y} + \textrm{H.c.}\right) \nonumber \\
	& - \sum_{x,y}  \left( \frac{t e^{-i\theta}}{2} \hat{a}_{2x,y+1}^\dagger \hat{a}_{2x,y} + \textrm{H.c.}\right),
\end{align}
where $\hat{a}^\dagger_{x,y}$ ($\hat{a}_{x,y}$) creates (annihilates) a boson at site $(x,y)$ of the primary lattice, $U$ is the onsite interaction strength, $J$ and $P$ represent the single-boson and two-boson hopping amplitudes along the $x$-direction, respectively, and $t$ is the hopping amplitude along the $y$-direction, which includes a nontrivial phase $\theta$. Note that the pair hopping has been both theoretically proposed \cite{PhysRevA.110.L020201, Eckholt2009} and experimentally \cite{PhysRevLett.111.173004, PhysRevLett.125.245301} observed in ultracold atomic systems.

The auxiliary lattice has a common decay rate $\lambda$, and its Hamiltonian is written as
\begin{align}\label{Hamilb}
	\mathcal{H}_\textrm{b}   =   -i\lambda \sum_{x,y} \hat{b}^\dagger_{x,y} \hat{b}_{x,y},
\end{align}
where $\hat{b}^\dagger_{x,y}$ ($\hat{b}_{x,y}$) is the bosonic creation (annihilation) operator at site ($x,y$) of the auxiliary lattice. 

The interaction between the dissipative primary and auxiliary lattices, with a coupling strength $g$, is given by
\begin{align}\label{Hamilint}
	\mathcal{H}_\textrm{int}   =  - g \sum_{x,y} \left(\hat{a}^\dagger_{x,y} \hat{b}_{x,y} + \hat{a}^\dagger_{x,y+1} \hat{b}_{x,y} + \textrm{H.c.}\right).
\end{align}
\begin{figure}[!b]
	\centering
	\includegraphics[width=8.7cm]{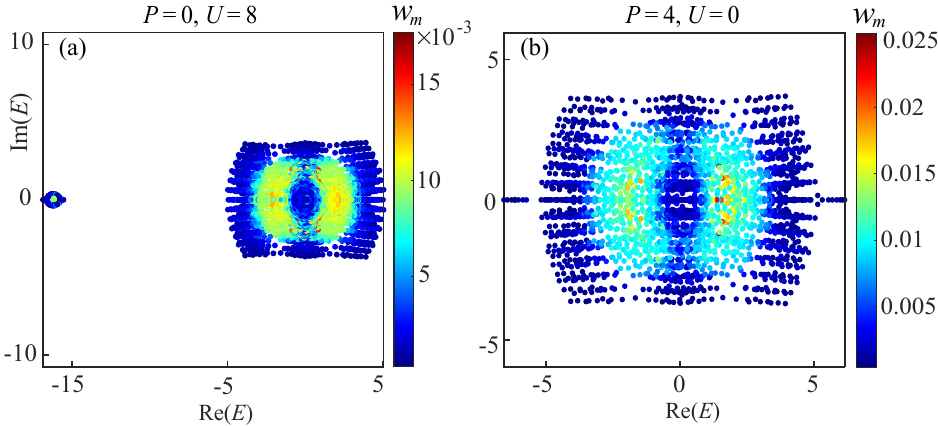}
	\caption{The complex eigenenergy spectrum, under OBCs along both the $x$- and $y$-directions, for (a) $P=0$ and (b) $U=0$. The color represents  the value of skin corner weight $w_{m}$. The other parameters are the same as those  in Fig.~\ref{Fig2}  with  $L_x \times L_y =9\times 8$.   }\label{FigS5}
\end{figure}

When the dissipation rate $\lambda$ is much larger than all the other energy scales, the fast-decaying modes in the auxiliary lattice can be adiabatically eliminated \cite{PhysRevA.86.012126,PhysRevA.85.032111,PhysRevX.13.031009}. This procedure yields an effective description of the dynamics of the primary lattice, encapsulated in a master equation that governs the evolution of its degrees of freedom exclusively
\begin{align}\label{densityoperator}
	\partial_t \hat{\rho} = -i \left[\mathcal{H}_\textrm{a},\hat{\rho}\right] + \mathcal{D}_\textrm{a}[\hat{\rho}].
\end{align}
Here, the dissipator takes the form 
\begin{align}\label{dissipator}
	\mathcal{D}_\textrm{a}[\hat{\rho}] &= -  \{\mathcal{H}_\textrm{D},\hat{\rho}\} \nonumber \\
	&+ \Gamma \sum_{x,y}\left(\hat{a}_{x,y+1} \hat{\rho} \hat{a}^\dagger_{x,y} + \hat{a}^\dagger_{x,y} \hat{\rho} \hat{a}_{x,y+1} + 2 \hat{a}_{x,y} \hat{\rho} \hat{a}^\dagger_{x,y}\right),
\end{align}
where $\mathcal{H}_\textrm{D} = \Gamma \sum_{x,y} \left[\left(\hat{a}^\dagger_{x,y} \hat{a}_{x,y+1} + \textrm{H.c.}\right)/2 + \hat{a}^\dagger_{x,y} \hat{a}_{x,y}\right]$ with the effective decay rate $\Gamma = 2 g^2/\lambda$. 

The quantum-trajectory theory states that the dynamics in Eqs.~(\ref{densityoperator}) and (\ref{dissipator}) can be  decomposed into nonunitary evolution and stochastic quantum jumps \cite{Meystre2007}.  By projecting out the quantum jumps by continuously monitoring the particle number \cite{PhysRevA.94.053615, PhysRevLett.124.147203}, the system's dynamics is governed by the effective non-Hermitian Hamiltonian     $\mathcal{H}_\textrm{NH} = \mathcal{H}_\textrm{a} - i \mathcal{H}_\textrm{D}$ as
\begin{align}\label{HamilNH}
	\mathcal{H}_\textrm{NH} =  & - J   \sum_{x,y}  \left( \hat{a}_{x+1,y}^\dagger  \hat{a}_{x,y} +\textrm{H.c.} \right)   \nonumber \\
	& - U \sum_{x,y} \hat{n}_{x,y}(\hat{n}_{x,y} - 1) -  it \sum_{x,y} \hat{n}_{x,y} \nonumber \\
	& - it \sum_{x,y}  \left( \hat{a}_{2x-1,y+1}^\dagger  \hat{a}_{2x-1,y} +  \hat{a}_{2x,y}^\dagger \hat{a}_{2x,y+1}\right)  \nonumber \\
	& -\frac{P}{2} \sum_{x,y} \left(\hat{a}_{2x-1,y}^\dagger \hat{a}_{2x-1,y}^\dagger  \hat{a}_{2x,y}  \hat{a}_{2x,y}+ \textrm{H.c.}\right),
\end{align}
where we choose the phase $\theta = \pi/2$,  and set the effective decay rate $\Gamma= t$ to realize unidirectional hopping along the $y$ direction. 
Without loss of generality, the total onsite dissipative amplitude in Eq.~(\ref{HamilNH}) can be neglected and treated as a background contribution. This leads us to the model Hamiltonian as
\begin{align}\label{hamilappendix}
	\mathcal{H}   =  & - J   \sum_{x,y}  \left( \hat{a}_{x+1,y}^\dagger  \hat{a}_{x,y} +\textrm{H.c.} \right) - U \sum_{x,y} \hat{n}_{x,y} \left(\hat{n}_{x,y}-1 \right) \nonumber \\
	& -i t \sum_{x,y}  \left(\hat{a}_{2x-1,y+1}^\dagger  \hat{a}_{2x-1,y} + \hat{a}_{2x,y}^\dagger  \hat{a}_{2x,y+1}\right) \nonumber \\
	& -\frac{P}{2} \sum_{x,y} \left(\hat{a}_{2x-1,y}^\dagger \hat{a}_{2x-1,y}^\dagger  \hat{a}_{2x,y}  \hat{a}_{2x,y}+ \textrm{H.c.}\right).
\end{align}
\begin{figure}[!b]
	\centering
	\includegraphics[width=8.8cm]{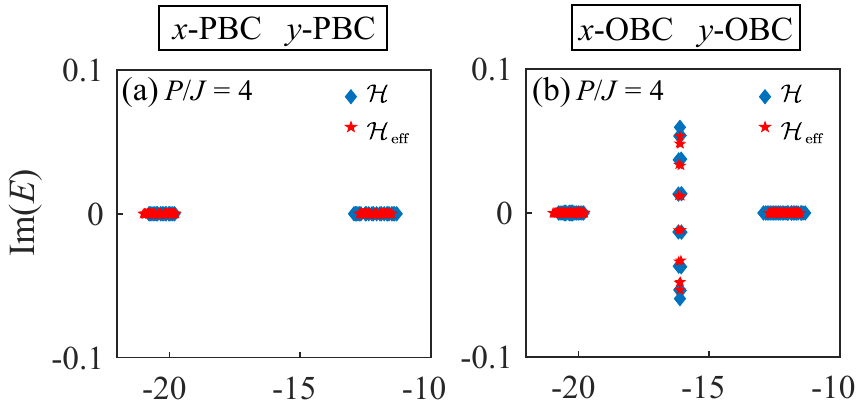}
	\caption{The complex eigenenergy spectrum of  doublon bands, calculated using the non-Hermitian interacting Hamiltonian $\mathcal{H}$ (blue diamonds) and effective Hamiltonian $\mathcal{H}_\textrm{eff}$ (red stars) under PBCs (a) and OBCs (b) along both the $x$- and $y$-directions. The parameters are the same as those  in Fig.~\ref{Fig2}  with  $L_x \times L_y =13\times 14$.   }\label{FigS2}
\end{figure}

\begin{figure*}[!bt]
	\centering
	\includegraphics[width=18cm]{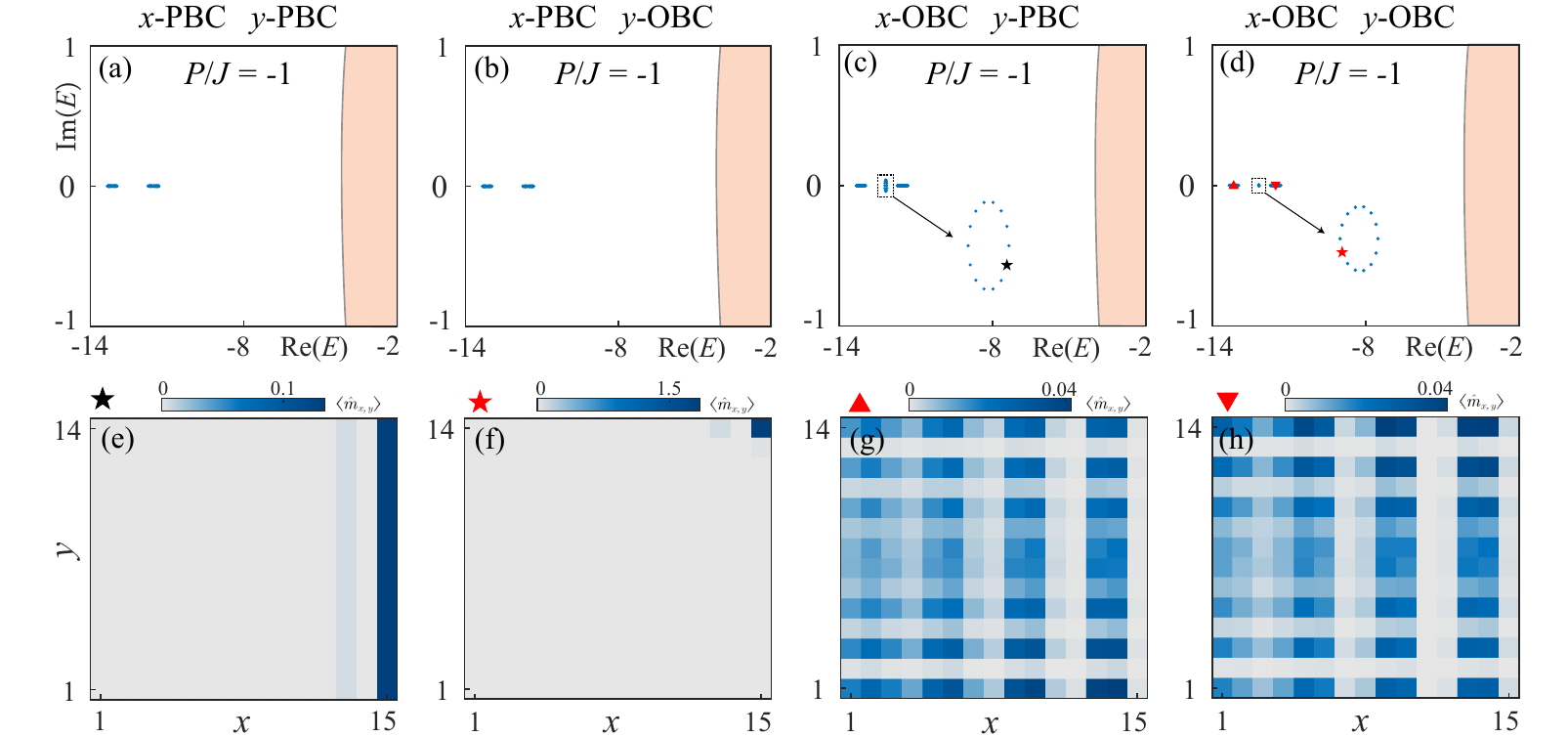}
	\caption{ (a-d) Complex eigenenergy spectrum of $\mathcal{H}$  for  $P/J=-1$ under various boundary conditions. The OBCs are imported along none of the directions in (a), only along the $y$-direction in (b), only along the $x$-direction in (c), and along both  directions in (d). Apricot-shaded areas represent the continuum of scattering states, and blue dots denote doublon states. The enlarged view of the dotted square region is highlighted by the arrows in (c,d). (g,h) The corresponding site-resolved   double-occupation boson density $\langle \hat{m}_{x,y} \rangle $ of the selected energies. The selected energies are marked by the black and red identifiers within the eigenenergy spectrum in (c) and (d). The lattice size  is set as $L_x \times L_y=15 \times 14$   for the open boundary along the $x$-direction, and for the other boundary conditions, it is $L_x \times L_y = 14 \times 14$. Other parameters are given as $t/J=0.5$ and $U/J=6$.}\label{Fig3}
\end{figure*}

\begin{figure}[!tb]
	\centering
	\includegraphics[width=8.7cm]{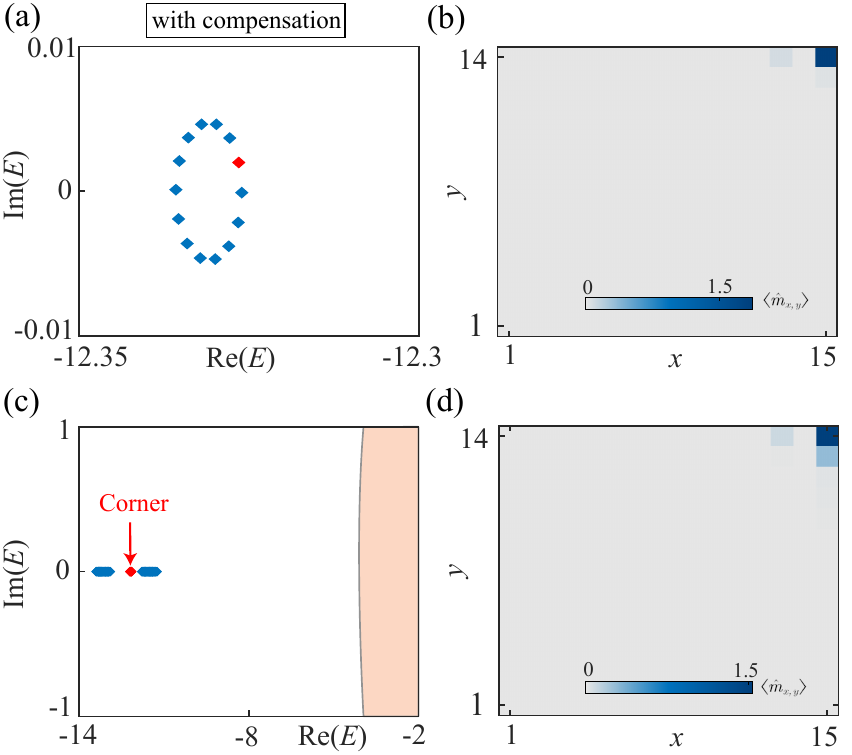}
	\caption{ (a) Complex eigenenergy spectrum of in-gap corner skin modes  states   with the potential $V = J^2/U$ applied   to both left and right edges to compensate for the interaction-induced detuning.  The corresponding site-resolved double-occupation density $\langle \hat{m}_{x,y} \rangle$ for  the selected in-gap state [see red marker in (a)] is plotted in (b). (c) Complex eigenenergy spectrum of $\mathcal{H}$, subject to the random disorder applied to the hopping terms $J$, $t$ and $P$  for  $P/J=-1$ and $W/J=0.3$. (d) The corresponding site-resolved density $\langle \hat{m}_{x,y} \rangle$ of the selected corner state.  The parameters are the same as those in Fig.~\ref{Fig3}.}\label{FigS4}
\end{figure}

\section{Eigenspectrum for $U=0$ or $P=0$}\label{AppendixB2}

The system Hamiltonian $\mathcal{H} $ in Eq.~(\ref{hamil}) exhibits no corner skin modes in the absence of many-body interactions $U$ or $P$. Here, we calculate its eigenspectrum for $U=0$ or $P=0$.

To check whether the system supports corner skin modes for  $U=0$ or $P=0$, we calculate the state-dependent skin corner weight, which is  defined as
\begin{align}\label{cornerskin}
	w_{m} = \frac{1}{2}\sum_{ \, r,\, r_c} \abs{\frac{_R\langle \psi_m | \hat{n}_{r} | \psi_m \rangle_R}{_R \langle \psi_m | \psi_m \rangle_R}}^2 \exp\left(-\frac{|r - r_c|}{\xi}\right),
\end{align}
where the sum runs over all the lattice sites $r=(x,y)$ and corner positions \(r_c\), and $\xi=1$ is the decay length. When $w_{m}$ is close to one, it indicates the existence of well-localized corner modes.  

Figure \ref{FigS5} shows the  complex eigenenergy spectrum for $U=0$ or $P=0$  under OBCs along both the $x$- and $y$-directions. The color represents  the value of skin corner weight $w_{m}$. A small value of $w_{m}$ indicates the absence of corner skin modes for $U=0$ or $P=0$.

\section{Effective Doublon Hamiltonian in Strong-Interaction Limit}\label{AppendixB}

To gain an intuitive understanding of the mechanism of interaction-induced second-order skin effects underlying the two-body interaction, we derive the effective Hamiltonian for two-body bound states (doublons) in the strong-interaction limit. 

We begin by expressing the Hamiltonian in Eq.~(\ref{hamilappendix}) as $\mathcal{H} = \mathcal{H}_\textrm{0} + \mathcal{V}$, where $\mathcal{H}_\textrm{0}$ is the dominant interaction part with
\begin{align}\label{hamil00}
	\mathcal{H}_\textrm{0}   = - U \sum_{x,y} \hat{n}_{x,y} \left(\hat{n}_{x,y}-1 \right) ,
\end{align}
and the hopping terms $\mathcal{V}$ reads 
\begin{align}\label{V}
	\mathcal{V}   =&  - J   \sum_{x,y}  \left( \hat{a}_{x+1,y}^\dagger  \hat{a}_{x,y} +\textrm{H.c.} \right) \notag \nonumber \\
	&-i t \sum_{x,y}  \left(\hat{a}_{2x-1,y+1}^\dagger  \hat{a}_{2x-1,y} + \hat{a}_{2x,y}^\dagger  \hat{a}_{2x,y+1}\right) \notag \nonumber \\
	&-\frac{P}{2} \sum_{x,y} \left(\hat{a}_{2x-1,y}^\dagger \hat{a}_{2x-1,y}^\dagger  \hat{a}_{2x,y}  \hat{a}_{2x,y} +  \textrm{H.c.}\right).
\end{align}
In the strong-interaction limit with $\abs{U} \gg ~\abs{J}, ~\abs{t},~\abs{P}$, we treat the hopping term $\mathcal{V}$ in Eq.~(\ref{V}) as a perturbation. In this regime, two bosons are tightly bound to form a bound state (doublon), such that the set of doublon states $|d_{x,y}\rangle = \hat{d}^\dagger_{x,y} |0\rangle= \hat{a}^\dagger_{x,y} \hat{a}^\dagger_{x,y} |0\rangle$ are eigenstates of the unperturbed Hamiltonian $\mathcal{H}_\textrm{0}$ with energy $E_d = -2U$. All the other two-boson scattering states are of form $|s\rangle = \hat{a}^\dagger_{x',y'} \hat{a}^\dagger_{x,y} |0\rangle$, where $x'\neq x,~y'\neq y$, and have energy $E_s = 0$. For convenience, we adopt the shorthand notation $\ket{d}\equiv\ket{d_{x,y}}$ in the following.

We now proceed to derive the effective Hamiltonian $\mathcal{H}_\textrm{eff}$ by utilizing quasi-degenerate second-order perturbation theory \cite{Bir1974,CCohenTannoudji1Atom,PhysRevResearch.2.013348,PhysRevA.97.013637}. The nonzero matrix elements of the effective Hamiltonian are given by 
\begin{align}\label{perturbation}
	\langle d|\mathcal{H}_\textrm{eff}|d'\rangle = & ~E_d \delta_{d,d'} + \langle d |\mathcal{V}|d'\rangle \nonumber \\
	&+  \frac{1}{2} \sum_{s} \langle d|\mathcal{V}|s\rangle \langle s |\mathcal{V}|d'\rangle \times \left(\frac{2}{E_d-E_s}\right),
\end{align}
where $E_n$ is the $n$th eigenvalue  of $\mathcal{H}_\textrm{0}$ relative to eigenstate $|n\rangle$. When $|d\rangle = |d'\rangle$, the matrix element in Eq.~(\ref{perturbation}) yields the effective doublon offsite energy $-U_\textrm{eff} \equiv \langle d|\mathcal{H}_\textrm{eff}|d\rangle$. In the case where $|d\rangle \neq |d'\rangle$, these matrix elements describe the effective doublon hopping $-J_\textrm{eff}/t_\text{eff} \equiv \langle d|\mathcal{H}_\textrm{eff}|d'\rangle$ along the $x$  or $y$ direction. 

By inserting Eqs.~(\ref{hamil00}-\ref{V}) into Eq.~(\ref{perturbation}), we obtain the effective Hamiltonian acted on quasi-single-particle subspace as
\begin{align}\label{hamileffappendix}
	\mathcal{H}_\textrm{eff} = & - U_\textrm{eff} \sum_{x,y} \hat{d}^\dagger_{x,y} \hat{d}_{x,y} - J_\text{eff} \sum_{x,y} \left( \hat{d}^\dagger_{2x+1,y} \hat{d}_{2x,y} + \text{H.c.}\right) \notag \nonumber \\
	&- (J_\text{eff}+P) \sum_{x,y} \left(\hat{d}^\dagger_{2x,y} \hat{d}_{2x-1,y} + \text{H.c.}\right) \notag \nonumber \\
	&+ t_\text{eff} \sum_{x,y} \left(\hat{d}^\dagger_{2x-1,y+1} \hat{d}_{2x-1,y} + \hat{d}^\dagger_{2x,y} \hat{d}_{2x,y+1}\right),
\end{align}
where $J_\textrm{eff} = J^2/U$ is the effective symmetric doublon hopping along the $x$ direction, and $t_\textrm{eff} = t^2/U$ is the effective unidirectional along the $y$ direction. For the PBCs, the effective onsite energy is given by $U_\textrm{eff} = 2J^2/U+2U$. When the boundary along the $x$ direction is open, the onsite energy reads $U_\textrm{eff} = 2J^2/U+2U$ for the bulk sites, and $U_\textrm{eff} = J^2/U+2U$ for sites located at the left and right edges.
 
\begin{figure}[!b]
	\centering
	\includegraphics[width=8.7cm]{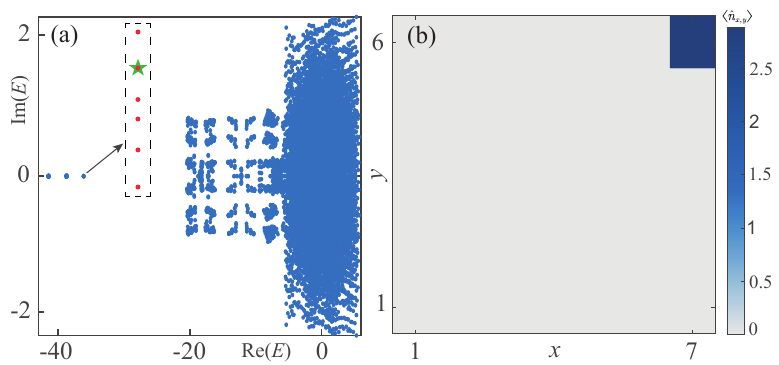}
	\caption{ (a) Complex eigenenergy spectrum of $\mathcal{H}$ with three-excitation subspace under OBCs along both the $x$ and $y$ directions. The enlarged view highlights the corner skin modes in red dots. (b) The corresponding site-resolved single-occupation boson density $\langle \hat{n}_{x,y} \rangle$   for the selected energy, marked by a green star in the eigenenergy spectrum  (a).  Parameters used are $P/J=6$, $t/J=1$, and $U/J=6$ with $L_x \times L_y=7 \times 6$.  }\label{FigS6}
\end{figure}

In the strong-interaction limit, as described by Eq.~(\ref{hamileffappendix}), a doublon behaves as a quasiparticle within a Hermitian Su–Schrieffer–Heeger (SSH) chain in the $x$ direction. This chain exhibits staggered hopping amplitudes of $J_\textrm{eff} + P$ and $J_\textrm{eff}$. However, unlike standard SSH model \cite{PhysRevLett.42.1698},   onsite energies at edge sites in the Hamiltonian $\mathcal{H}$   are detuned by $J^2/U$ due to onsite many-body interactions. Along the $y$-direction, these SSH chains are coupled through an alternating unidirectional hopping denoted by $t_\text{eff}$.  

Figure~\ref{FigS2} illustrates the energy spectra of doublon under both OBCs and PBCs, calculated using the non-Hermitian interacting Hamiltonian $\mathcal{H}$ (blue diamonds) and the effective Hamiltonian $\mathcal{H}_\textrm{eff}$ (red stars). In the strong-interaction limit, the energy spectra of the two Hamiltonians align closely, showing excellent agreement.

\section{Corner skin modes for $P<-2J^2/U$}\label{AppendixD}

According to Eq.~(\ref{condition}), corner skin modes can appear for  $P>0$ or $P < -2J^2/U$.  While we have discussed the case of $P>0$ in the main text,  here we present the results for $P < -2J^2/U$.

Figure \ref{Fig3} shows the complex eigenspectrum  $E$ and the corresponding site-resolved  double-occupation boson density $\langle \hat{m}_{x,y} \rangle $ for $P/J=-1$ under different boundary conditions,  where the apricot-shaded regions represent the scattering states, and the blue dots correspond to doublon states. The eigenspectrum is distinctly partitioned into a continuum of scattering states and discrete doublon bands.  As in the case of  $P/J=4$, a band gap separates the two doublon bands  under PBCs along both the $x$ and $y$ directions, as shown in Fig.~\ref{Fig3}(a). 

When the boundary condition  only along  the $y$ direction is open, no in-gap modes are observed [see Fig.~\ref{Fig3}(b)]. In contrast, when the boundary is open only along the $x$ direction, in-gap modes emerge [see Figs.~\ref{Fig3}(c)]. These in-gap modes are localized at the right edge, as illustrated in Fig.~\ref{Fig3}(e), which can be understood as topological weakly-linked edge modes [see details in  Sec.~III(B)].  In the case of OBCs along both the $x$ and $y$ directions, in-gap modes still persist [see Fig.~\ref{Fig3}(d)].  Due to the unidirectional hopping along the $y$ direction in the right boundary,  these in-gap modes become localized at the top-right corners [see Fig.~\ref{Fig3}(f)]. In contrast, the bulk states of doublons remain extended across the bulk sites due to destructive interference caused by the opposing unidirectional hopping along the $y$ directions within each square plaquette [see   Fig.~\ref{Fig3}(g,h)].

To verify that the corner states discussed above are induced by the interplay between the interaction-driven  topological edge states and the nonreciprocal hopping along the edge, rather than by Tamm-Shockley states, we introduce a compensating potential $V$, applied to the left and right edges. Figure   \ref{FigS4}(a)   presents  the   complex eigenenergy spectra of in-gap corner skin states for $P/J=-1$,   with the application of a compensating potential $V=J^2/U$.   The corresponding site-resolved   density $\langle \hat{n}_{x,y} \rangle$ for   the specific in-gap state is shown in Fig.~\ref{FigS4}(b). After applying the compensated potential, the in-gap doublon states remain well localized at the corners. This indicates that the corner skin modes do not arise from Tamm-Shockley edge states but are instead induced by the interplay between the interaction-driven  topological edge states and the nonreciprocal hopping along the edge.

In addition, we calculate the complex eigenenergy spectrum of $\mathcal{H}$ under random disorder applied to  the hopping terms $J$, $t$ and $P$ for $P/J=-1$, as shown in Fig.~\ref{FigS4}(c). Despite the presence of strong disorder, the band gap of the doublon bands and the in-gap states persist. Moreover, these in-gap doublon states remain well localized at the top-right corner of the lattice [see Fig.~\ref{FigS4}(d)], demonstrating their robustness against disorder.

	\section{Corner skin modes in the three-excitation subspace}\label{AppendixE}

In addition to the double-excitation subspace, corner skin modes can also emerge in higher-excitation subspace. The second-order skin effect typically arises from the interplay between one-dimensional edge modes and local nonreciprocity \cite{PhysRevLett.123.016805,PhysRevB.102.205118,PhysRevLett.131.116601}. If the decoupled pair-hopping chain along the $x$ direction supports edge states under OBCs, then these edge states become localized toward the corners due to local unidirectional hopping when the chains are coupled.

In the strong-interaction limit, bosons tend to occupy the same sites. By applying quasi-degenerate second-order perturbation theory, as given in Eq.~(\ref{perturbation}), the effective Hamiltonian in the higher-excitation subspace can exhibit a structure similar to the SSH lattice found in the two-excitation subspace [see Eq.~(\ref{hamileffappendix})], supporting edge states.

Figure \ref{FigS6}(a) shows the complex eigenenergy spectrum of $\mathcal{H}$ in the three-excitation subspace under OBCs along both the $x$ and $y$ directions. The three bands with lower real parts correspond to states where bosons predominantly occupy the same sites, due to strong interactions. The isolated bands, highlighted by red dots, are localized at the top-right corner, indicating the presence of second-order skin effects in the three-excitation subspace [see the density distribution in Fig.~\ref{FigS6}(b)].

%

\end{document}